\documentclass[11pt,graphicx,amsmath]{article}
\usepackage{amsmath}
\usepackage{graphicx}
\usepackage{bm}
\usepackage{xcolor}
\usepackage{amssymb}
\usepackage{amsfonts}
\usepackage{comment}
\usepackage{cite}
\usepackage{todonotes}
\usepackage{caption}
\usepackage{subcaption}
\usepackage{CJKutf8}
\usepackage{authblk}

\usepackage[utf8]{inputenc}

\def\be{\begin{equation}}
\def\ee{\end{equation}}
\def\ba{\begin{eqnarray}}
\def\ea{\end{eqnarray}}
\def\nn{\nonumber}
\def\bl#1\el{\begin{align}#1\end{align}}

\def\be{\begin{equation}}
\def\ee{\end{equation}}
\def\ba{\begin{eqnarray}}
\def\ea{\end{eqnarray}}
\def\nn{\nonumber}

\def\bl#1\el{\begin{align}#1\end{align}}

\title{Regularized vacuum stress tensor of a scalar field as the inflaton or dark energy}

\author{Xuan Ye\thanks{ yexuan@cuhk.edu.cn}}
\affil{\small School of Science and Engineering, The Chinese University of Hong Kong (Shenzhen), Longgang, Shenzhen, Guangdong 518172, P.R. China}

 \date{}

 \date{}

\large

\begin{document}

\maketitle
\allowdisplaybreaks

\begin{abstract}

We study the regularized vacuum stress tensor of scalar fields in maximally symmetric spacetime and assess the feasibility of driving primordial inflation or current cosmic acceleration by analyzing the existence of solutions to the Friedmann equation. We find that a conformally coupled scalar field with mass of order $10$ $M_{\text{pl}}$ can be a candidate for both the inflaton and dark energy, suggesting that these two components may have the same quantum origin. In contrast, a minimally coupled scalar field cannot serve as either the inflaton or dark energy regardless of its mass.



\end{abstract}

\section{Introduction}\label{introduction}

The Universe has experienced two periods of quasi-exponential expansion, with an early inflationary phase having a Hubble rate 
$H_{\text{Inf}}\approx 10^{14}$ GeV \cite{Liddle1993} and the current accelerated expansion having a Hubble rate $H_{0}\approx 10^{-42}$ GeV \cite{Planck2018}. The inflationary epoch is strongly supported by observations, as it resolves the horizon and flatness problems of the standard Big Bang model and predicts a nearly scale-invariant primordial power spectrum \cite{Guth1981, Lindde1982, Weinberg2008, KalloshLinde2025, Bamba024, Planck2018}. The present accelerated expansion is confirmed by observations of Type Ia supernovae \cite{Riess1998, Perlmutter1999, Suzuki2012}, the cosmic microwave background \cite{Hinshaw2013, Planck2018}, and large-scale structure \cite{Alam2021}. Despite these successes, the physical mechanisms driving these two expansion phases remain open questions.

To realize an inflationary phase, most models invoke inflaton potentials with very small slopes or plateau-like regions at the onset of inflation, as in Starobinsky inflation, Higgs inflation, and $\alpha$-attractor models \cite{Starobinsky1980,Bezrukov2008,KalloshLinde2013}. In this regime, the stress tensor of the inflaton is dominated by its potential rather than its kinetic term, leading to an equation of state $w\approx-1$ and an approximately constant Hubble rate \cite{KalloshLinde2025, Bamba024}. To account for the present accelerated expansion, a wide range of dynamical field models \cite{Caldwell1998,Caldwell2002,ArmendarizPicon2001,Zheng2022,Zhang2007,Xia2007,Wang2008a,Wang2008b,Zhao2009,Don2016,Bose2025} and alternative theories of gravity \cite{Carroll2004,Nicolis2009,deRham2011,Bengochea2009} have been proposed. Besides these non-trivial models, a simple cosmological constant term $\Lambda g_{\mu\nu}$ added to the Einstein equation can also explain the accelerated expansion if it contributes about $70\%$ of the critical energy density \cite{Weinberg2008,Maggiore2018,Mukhanov2005}, known as the standard $\Lambda$CDM model. In this case, 
the equation of state is $w=-1$. A flat, homogeneous, and isotropic spacetime dominated by a cosmological constant undergoes exponential expansion, allowing the cosmological constant model to phenomenologically describe both inflation \cite{LiddleLyth2000} and present accelerated expansion \cite{Dodelson2020}. Nevertheless, the origin of the cosmological constant remains unclear \cite{weinberg1989}.

In this paper, we study the feasibility of the vacuum stress tensor of a minimally or conformally coupled scalar field can act as the inflaton, dark energy, or both. For the inflationary phase, unlike conventional models, we do not introduce a specific potential. Instead, we ask whether vacuum fluctuations of a scalar field alone can drive inflation and estimate the parameter values required to sustain it. 
We assume that the scalar field initially resides in a maximally symmetric spacetime, motivated by the extremely high energy density at the onset of inflation, of order $(H_{\text{Inf}})^4 \approx  10^{56}$ GeV$^{4}$. For the present accelerated expansion, the Universe is homogeneous and isotropic on large scales and undergoes quasi-exponential expansion with the energy density $(H_{0})^4\approx 10^{-168}$ GeV$^{4}$, dominated by dark energy. The energy densities associated with these two epochs differ by many orders of magnitude, and whether they share a common origin has been discussed from different perspectives in the literature \cite{Kaneta,Joan,Rubio}. Here, we address the question within the framework of the vacuum stress tensor of a scalar field with curvature coupling.

The vacuum stress tensor of quantum field in curved spacetime typically exhibits ultraviolet (UV) divergences. Therefore, before discussing any physical effects, these divergences must be properly regularized. In previous works, we studied regularization schemes for subtracting UV divergences of vacuum stress tensors in de Sitter space \cite{ZhangYe2025b,YeZhang2025,YeZhang2024,ZhangYe2022,XuanJCAP2022,ZhangYeWang2020,ZhangWangYe2020}. Following the minimal subtraction rule \cite{ParkerFulling1974}, the required regularization order depends on the field type, spacetime geometry, and field parameters. For example, for a conformally (minimally) coupled scalar field, regularization up to the 0th (2nd) adiabatic order is sufficient to remove UV divergences.
Furthermore, we have shown that the regularized vacuum stress tensors of massive scalars, spinors, and vector fields are constant and maximally symmetric, and can jointly contribute to a cosmological constant   \cite{XuanJCAP2022,ZhangYeWang2020,YeZhang2024,ZhangYe2022,YeZhang2025,ZhangYe2025b}. An important question is whether such contributions can act as the inflaton in the early universe or as dark energy driving the present accelerated expansion. Here, we address this issue by investigating the existence of viable solutions to the Friedmann equation. As a first step, we consider massive conformally and minimally coupled scalar fields with analytic expressions of the regularized vacuum stress tensor available from our previous work \cite{XuanJCAP2022,Ye2023}.

We then review the derivation of finite  vacuum stress tensors obtained using the 0th and 2nd order point-splitting regularization for conformally and minimally coupled scalar fields, respectively, and discuss whether these fields can account for inflation or dark energy.

\section{Point-splitting Regularization for Vacuum Stress Tensor}

In this section, we review the point-splitting method for deriving the UV divergent vacuum stress tensor, present the subtraction terms constructed from the adiabatic mode functions, and show how these terms remove the UV divergences \cite{XuanJCAP2022}.

\subsection{Unregularized vacuum stress tensor}

Consider a free scalar field $\phi$ propagating in a curved spacetime and obeying the curvature coupled Klein--Gordon equation,
\be
(\Box + m^2 + \xi R)\phi = 0,
\label{fieldequxi}
\ee
where $\Box \equiv \nabla_\alpha \nabla^\alpha$ is the d'Alembert operator, $R$ is the Ricci scalar, $m$ is the mass of the field, and $\xi$ is the coupling constant. The field is conformally coupled for $\xi = 1/6$ and minimally coupled for $\xi = 0$.

In a spatially flat Robertson--Walker spacetime with metric
\[
ds^2 = a(\tau)^2 (d\tau^2 - \delta_{ij} dx^i dx^j),
\]
the field operator can be expanded in Fourier modes as
\be
\phi(\mathbf{x},\tau) = \int \frac{d^3 k}{(2\pi)^{3/2}}
\left[
a_{\mathbf{k}} \phi_k(\tau) e^{i \mathbf{k}\cdot \mathbf{x}}
+ a^\dagger_{\mathbf{k}} \phi_k^*(\tau) e^{-i \mathbf{k}\cdot \mathbf{x}}
\right],
\label{operexation}
\ee
where $\tau$ is conformal time, and $a_{\mathbf{k}}$ and $a^\dagger_{\mathbf{k}}$ are annihilation and creation operators satisfying the canonical commutation relation
$[a_{\mathbf{k}}, a^\dagger_{\mathbf{k}'}] = \delta(\mathbf{k} - \mathbf{k}')$.

In de Sitter space the scale factor is
\be
a(\tau) = -\frac{1}{H\tau},
\label{scalefactor}
\ee
where $H$ is a constant Hubble rate. Although the inflationary and current accelerating regimes undergo quasi-exponential expansion, they can be roughly described by the scale factor \eqref{scalefactor}. Various inflation models have different expansion rates. For specificity, we take  \cite{Liddle1993}
\be
H_{\mathrm{Inf}} \approx 10^{14}\,\mathrm{GeV} \approx 10^{-5} M_{\mathrm{pl}}.
\ee
For the present accelerating expansion, we use \cite{Planck2018}
\be
H_0 \approx 10^{-42}\,\mathrm{GeV} \approx 10^{-61} M_{\mathrm{pl}},
\ee
where $M_{\mathrm{pl}} = G^{-1/2}$ is the Planck mass.

Defining $v_k \equiv a \phi_k$, the mode equation is
\be
v_k'' + \left[
k^2 + m^2 a^2 + \left(\xi - \frac{1}{6}\right) a^2 R
\right] v_k = 0,
\label{equxi}
\ee
where a prime denotes differentiation with respect to conformal time. The Ricci scalar is given by $R = 6 a'' / a^3$, which reduces to $R = 12 H^2$ in de Sitter space. The positive frequency solution of Eq.~\eqref{equxi} in de Sitter space is
\be  
v_k (\tau )  =  \sqrt{\frac{\pi}{2}}\sqrt{\frac{x}{2k}}
  e^{i \frac{\pi}{2}(\nu+ \frac12) } H^{(1)}_{\nu} ( x) ,\label{u}
\ee
where $H^{(1)}_\nu$ is the Hankel function of the first kind,
$x \equiv -k\tau$, and $\nu \equiv \sqrt{9/4 - m^2/H^2 - 12\xi }$.
In the high-$k$ limit, the solution \eqref{u} approaches the plane wave form
$(2k)^{-1/2} e^{-ik\tau}$, which corresponds to the Bunch--Davies vacuum state
$|0\rangle$ defined by $a_{\mathbf{k}}|0\rangle = 0$.

The unregularized Green's function in the Bunch--Davies vacuum is \cite{XuanJCAP2022}
\bl
G(\sigma) &\equiv \langle 0 | \phi(x^\mu)\phi(x'^{\mu}) | 0 \rangle
\label{GreeHyper}
\\
&= \frac{H^2}{16\pi^2}
\Gamma\!\left(\frac{3}{2}-\nu\right)
\Gamma\!\left(\nu+\frac{3}{2}\right)
\,{}_2F_1\!\left[
\frac{3}{2}+\nu, \frac{3}{2}-\nu, 2,
1+\frac{\sigma}{2}
\right],
\nn
\el
where the mode expansion \eqref{operexation} has been used.
Here ${}_2F_1$ is the hypergeometric function and $\Gamma$ denotes the Gamma
function. The variable $\sigma$ is defined as $\sigma \equiv (2\tau\tau')^{-1}
\big[ (\tau-\tau')^2 - |\mathbf{r}-\mathbf{r}'|^2 \big].$
As $\sigma \to 0$, the Green's function exhibits UV divergences and has the expansion
\bl
G(\sigma) \approx \frac{1}{16\pi^2}
\left(
-\frac{1}{\epsilon^{2}}
+ W \ln \epsilon^2
+ X
+ Y \epsilon^2 \ln \epsilon^2
+ Z \epsilon^2
\right),\label{highkexpansiuon}
\el
where $\epsilon^2 \equiv \sigma/H^2$, and $W,X,Y,Z$ are functions of $H$ and $m$
\cite{XuanJCAP2022}. The Green's function at the coincidence limit therefore contains quadratic and logarithmic divergences, as well as finite terms. The counterterms required to subtract these divergences can be constructed from the Fourier transform of the adiabatic power spectrum \cite{ZhangYeWang2020}.
Subtracting these counterterms from Eq.\eqref{GreeHyper} yields a UV and Infrared convergent regularized Green's function.
Since this paper focuses on the regularized stress tensor, we do not explicitly regularize the Green's function here.

The unregularized stress tensor of a scalar field in the point-splitting scheme is defined as
{\small
\bl
\langle T_{\mu\nu} \rangle
&\equiv \lim_{x'\to x} \mathcal{T}_{\mu\nu} G(\sigma),
\label{gnt}
\\
&=
\lim_{x'\to x}
\Big[
\frac{1}{2}(1-2\xi)
(\nabla_\mu\nabla_{\nu'} + \nabla_{\mu'}\nabla_\nu)-\xi(\nabla_\mu\nabla_\nu+\nabla_{\mu'}\nabla_{\nu'})+(2\xi-\tfrac12)
\nn
\\
&\quad
 g_{\mu\nu}\nabla_\sigma\nabla^{\sigma'}
+\xi g_{\mu\nu}
(\nabla_\sigma\nabla^\sigma+\nabla_{\sigma'}\nabla^{\sigma'})
-\xi G_{\mu\nu}
+\tfrac12 m^2 g_{\mu\nu}
\Big] G(\sigma),\nn
\el}
where $\mathcal{T}_{\mu\nu}$ is a differential operator and
$G_{\mu\nu}$ denotes the Einstein tensor.
Here $\nabla_\mu$ and $\nabla_{\mu'}$ are covariant derivatives with respect to
$x$ and $x'$, respectively. Substituting \eqref{GreeHyper} into \eqref{gnt} and taking the coincidence limit
$x'\to x$, one finds 
\bl
\langle T_{\mu\nu} \rangle
=&
\left(
\frac{1}{\epsilon^4},
\frac{1}{\epsilon^2},
\ln\epsilon^2
\right)
+ \frac{1}{2} g_{\mu\nu} Y \ln\epsilon\nn
\\
&+ \frac{1}{2} g_{\mu\nu}\frac{R}{12\epsilon^2}
+ \frac{1}{2} g_{\mu\nu}\frac{W}{\epsilon^2}
+ (Y+\tfrac12 Z) g_{\mu\nu}.
\label{unreg}
\el
The unregularized stress tensor therefore contains quartic, quadratic, and logarithmic divergences, together with finite terms.

\subsection{Regularized vacuum stress tensors} \label{unregv2}

The vacuum stress tensor \eqref{unreg} exhibits UV divergences in the coincidence limit. To remove these divergences while preserving the covariant conservation of the stress tensor, one introduces counterterms for the Green's function, denoted by $G_{\text{sub}}(\sigma)$. This function shares the same divergent structure as the unregularized Green's function \eqref{highkexpansiuon}, but differs in its convergent terms. To this end, we employ the adiabatic mode function
\bl
v_{k}^{(n)}=\frac{1}{\sqrt{2\sum_{i=0}^{n}\Omega_{k}^{(i)}}}
\exp\!\left[-i \int^{\tau} \sum_{i=0}^{n}\Omega_{k}^{(i)}(\tau')\, d\tau' \right],
\label{modefunctio1n}
\el
where $\Omega_{k}^{(i)}$ are the effective frequencies of adiabatic order $i$, indicating the number of time derivatives involved. By substituting \eqref{modefunctio1n} into \eqref{equxi} and solving iteratively order by order, one obtains \cite{ZhangYeWang2020}
\bl
\Omega_{k}^{(0)}&\equiv \omega = \sqrt{k^2+m^2 a^2},
\label{0thmodefunction}
\\
\Omega_{k}^{(2)}&=3\Big(\xi-\frac{1}{6}\Big)\frac{1}{\omega}\frac{a''}{a}
-\frac{m^2 a^2}{4\omega^3}
\Big(\frac{a''}{a}+\frac{a'^2}{a^2}\Big)
+\frac{5m^4 a^4}{8\omega^5}\frac{a'^2}{a^2}.
\label{2ndmodefunction}
\el
The adiabatic mode function \eqref{modefunctio1n} satisfies Eq.~\eqref{equxi} at each adiabatic order, thereby ensuring the conservation of the adiabatic stress tensor at each order. As the adiabatic order increases, the mode function approaches the exact solution \eqref{u} and becomes more accurate in the ultraviolet regime. By replacing the exact mode functions with the adiabatic ones in the definition of the Green's function \eqref{GreeHyper}, one obtains the 0th and 2nd order subtraction Green's functions
\be
G(\sigma)^{(0)}_{\text{sub}} = \frac{H^2}{4\pi^2}
\frac{m}{H}\frac{1}{\sqrt{-2\sigma}}
K_1\!\left(\frac{m}{H}\sqrt{-2\sigma}\right),
\label{grsbxi16}
\ee
and
{\small
\bl
G(\sigma)^{(2)}_{\text{sub}}
&= \frac{H^2}{4\pi^2}\Bigg[
\frac{m}{H}\frac{1}{\sqrt{-2\sigma}}
K_1\!\left(\frac{m}{H}\sqrt{-2\sigma}\right)
+(1-6\xi)
K_0\!\left(\frac{m}{H}\sqrt{-2\sigma}\right)
\nn
\\
&\quad
+\frac{1}{4}\frac{m}{H}\sqrt{-2\sigma}
K_1\!\left(\frac{m}{H}\sqrt{-2\sigma}\right)
+\frac{\sigma}{12}\frac{m^2}{H^2}
K_2\!\left(\frac{m}{H}\sqrt{-2\sigma}\right)
\Bigg],
\label{sungrxi0}
\el}
where $K_0$, $K_1$, and $K_2$ are modified Bessel functions of the second kind.

The 0th order regularized stress tensor for a conformally coupled scalar field is \cite{XuanJCAP2022}
\bl
\langle T_{\mu\nu}\rangle_{\text{reg}}
&\equiv
\lim_{x'\to x}\mathcal{T}_{\mu\nu}G(\sigma)
-\lim_{x'\to x}\mathcal{T}_{\mu\nu}G^{(0)}_{\text{sub}}(\sigma)
\nn
\\
&=
g_{\mu\nu}\frac{m^4}{64\pi^2}
\Big[
\psi\!\Big(\frac{3}{2}-\nu_c\Big)
+\psi\!\Big(\frac{3}{2}+\nu_c\Big)
+\ln\frac{R}{12m^2}
\Big],
\label{eqmc}
\el
where $\nu_c\equiv \sqrt{1/4-m^2/H^2}$. The 2nd order regularized stress tensor for a minimally coupled scalar field is \cite{XuanJCAP2022}
{\bl
\langle T_{\mu\nu}\rangle_{\text{reg}}
\equiv&
\lim_{x'\to x}\mathcal{T}_{\mu\nu}G(\sigma)
-\lim_{x'\to x}\mathcal{T}_{\mu\nu}G^{(2)}_{\text{sub}}(\sigma)
\nn
\\
=&
\frac{g_{\mu\nu}}{64\pi^2}
\Bigg[
m^2\Big(m^2-\frac{R}{6}\Big)
\Big(
\psi\!\Big(\frac{3}{2}-\nu_m\Big)
+\psi\!\Big(\frac{3}{2}+\nu_m\Big)\nn
\\
&+\ln\frac{R}{12m^2}
\Big)
+\frac{m^2R}{9}
+\frac{R^2}{24}
\Bigg],
\label{eqm}
\el
where $\nu_m\equiv\sqrt{9/4-m^2/H^2}$ and $\psi(x)$ is the digamma function.

\begin{figure}[htbp]
    \centering
    \includegraphics[width=0.6\textwidth]{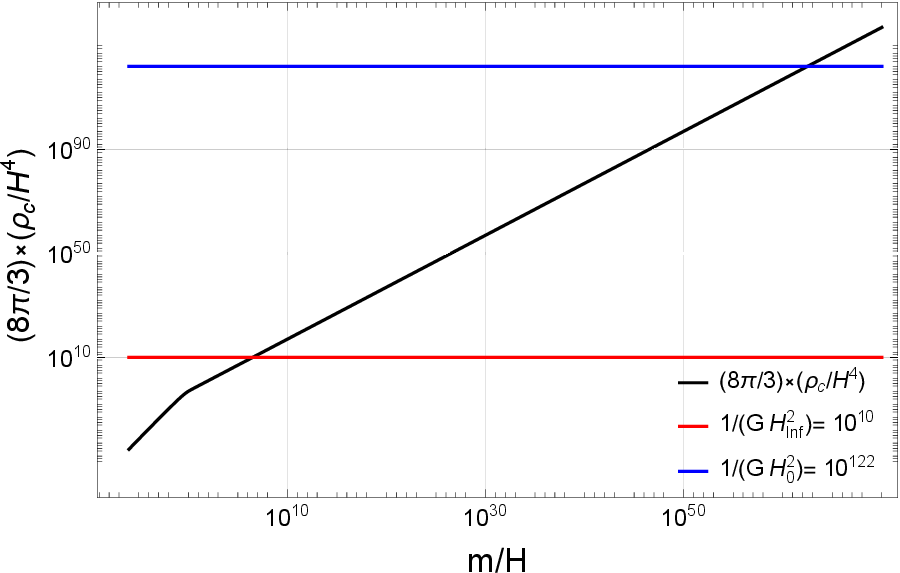}
\caption{Test of whether a conformally coupled scalar field can satisfy the Friedmann equation during inflation and the present accelerating expansion. Black line: $\frac{8\pi G}{3}\rho_c$. Red line: $1/(G H_{\text{Inf}}^2)$. Blue line: $1/(G H_{0}^2)$ }
    \label{Fig1}
\end{figure}

The stress tensors \eqref{eqmc} and \eqref{eqm} are both finite and maximally symmetric, and can therefore be interpreted as effective cosmological constant terms, consistent with our previous results \cite{XuanJCAP2022}. In the following sections, we further examine whether these vacuum stress tensors can drive primordial inflation and the present accelerating expansion.

\section{Vacuum Stress Tensor for the Conformally Coupled Scalar Field}

The regularized stress tensor \eqref{eqmc} of a conformally coupled scalar field is maximally symmetric in de Sitter space,
\bl
\langle T_{\mu\nu}\rangle_{\text{reg}}=
g_{\mu\nu}\Lambda_c,
\label{cctmunu}
\el
where the constant $\Lambda_c$ is defined as
\bl
\Lambda_c \equiv \frac{m^4}{64\pi^2}
\Big[
\psi\!\Big(\frac{3}{2}-\nu_c\Big)
+\psi\!\Big(\frac{3}{2}+\nu_c\Big)
+\ln\frac{H^2}{m^2}
\Big],
\label{eqiatop2c}
\el
and we have used the scale factor \eqref{scalefactor}. As shown in Fig.~\ref{Fig1}, the black curve shows that the regularized energy density
$\rho_c \equiv \Lambda_c$ is positive definite. Eq. \eqref{eqiatop2c} therefore represents an effective cosmological constant.

In the small-mass limit $(m/H)^2\to 0$, Eq.~\eqref{eqiatop2c} reduces to
\bl
\lim_{m^2/H^2\to 0}\Lambda_c
\approx
\frac{H^4}{64\pi^2}
\Big(\frac{m^2}{H^2}\Big)^2
\ln\frac{H^2}{m^2},
\label{masslesslimitc}
\el
which depends explicitly on both the mass and the Hubble rate. The cosmological constant \eqref{masslesslimitc} vanishes  in the limit $(m/H)\to 0$.

Taking the stress tensor \eqref{cctmunu} as the only source of the Einstein equations, the Friedmann equation reads
\bl
\Big(\frac{a'}{a}\Big)^2
=\frac{8\pi G a^2}{3}\rho_c
=\frac{G H^4 a^2}{24\pi}
\Big(\frac{m^2}{H^2}\Big)^2
\ln\frac{H^2}{m^2}.
\label{dehwjuifr}
\el
Substituting the scale factor \eqref{scalefactor} into Eq.~\eqref{dehwjuifr} gives
\bl
\frac{1}{G H^2}
=\frac{1}{24\pi}
\Big(\frac{m^2}{H^2}\Big)^2
\ln\frac{H^2}{m^2}.
\label{maindflation}
\el
The positivity of both sides of \eqref{maindflation} requires $m^2/H^2<1$.
The right hand side of Eq.~\eqref{maindflation} reaches a maximum value of
$2.9\times 10^{-4}$ at $(m/H)^2\approx 0.6$.
However,
this value is far smaller than the left hand side during inflation,
where $1/(G H_{\mathrm{Inf}}^2)\approx 10^{10}$,
and during the present accelerating expansion,
where $1/(G H_0^2)\approx 10^{122}$.
Eq.~\eqref{maindflation} has no solution in either epoch, as shown in Fig.1 at the low mass regime. Therefore, 
a conformally coupled scalar field in the small mass regime
$(m/H)^2\ll 1$
cannot act as the source of either inflation or dark energy.

In the large mass limit $(m/H)^2\to\infty$,
Eq.~\eqref{eqiatop2c} reduces to
\bl
\lim_{m^2/H^2\to\infty}\Lambda_c
\approx
\frac{H^4}{64\pi^2}
\frac{2}{3}\frac{m^2}{H^2},
\label{massivelicondtc}
\el
which depends on both the mass and the Hubble rate.
Using \eqref{cctmunu}, as the source of the Friedmann equation then gives
\bl
\Big(\frac{a'}{a}\Big)^2
=\frac{8\pi G a^2}{3}\rho_c
=\frac{G H^4 a^2}{36\pi^2}\frac{m^2}{H^2},
\label{dewjio}
\el
and substituting the scale factor \eqref{scalefactor} yields
\bl
m = \sqrt{36\pi^2} G^{-1/2}
\approx 10 M_{\mathrm{pl}}\approx 10^{20}\,\mathrm{GeV},
\label{dehwiudeh}
\el
which is independent of the Hubble rate.
This means that \eqref{dehwiudeh}  can apply to both the inflation and accelerating expansion. The ratios of the mass to Hubble rates are 
\bl
m/H_{\mathrm{Inf}} &\approx 10^6 \gg 1,
\label{Hpl1}
\\
m/H_0 &\approx 10^{62} \gg 1,
\label{Hpl0}
\el
which are consistent with the large mass assumption. These values correspond to the intersections of
$1/(G H_{\mathrm{Inf}}^2)$ and
$1/(G H_0^2)$
with $\tfrac{8\pi G}{3}\rho_c$, respectively, see Fig.\ref{Fig1}.
These results indicate that a supermassive conformally coupled scalar field with
can serve as a candidate for both the inflaton and dark energy,
driving the early inflationary expansion and the present accelerating expansion.
Because the field is extremely massive, it cannot be excited into particle states,
and only its vacuum effects can have observable consequences.

\section{Vacuum Stress Tensor for Minimally Coupled Scalar Field}

The regularized stress tensor \eqref{eqm} of a minimally coupled scalar field
is maximally symmetric in de Sitter space,
\bl
\langle T_{\mu\nu}\rangle_{\text{reg}}
= g_{\mu\nu}\Lambda_m,
\label{mctmunu}
\el
where the constant $\Lambda_m$ is defined as
\bl
\Lambda_m \equiv &\frac{H^4}{64\pi^2}
\Bigg[
\frac{m^2}{H^2}\Big(\frac{m^2}{H^2}-2\Big)
\Big(
\psi\!\Big(\frac{3}{2}-\nu_m\Big)
+\psi\!\Big(\frac{3}{2}+\nu_m\Big)\nn
\\
&+\ln\frac{H^2}{m^2}
\Big)
+\frac{4}{3}\frac{m^2}{H^2}
+6
\Bigg].
\label{eqiatop2}
\el

\begin{figure}[htbp]
    \centering
    \includegraphics[width=0.6\textwidth]{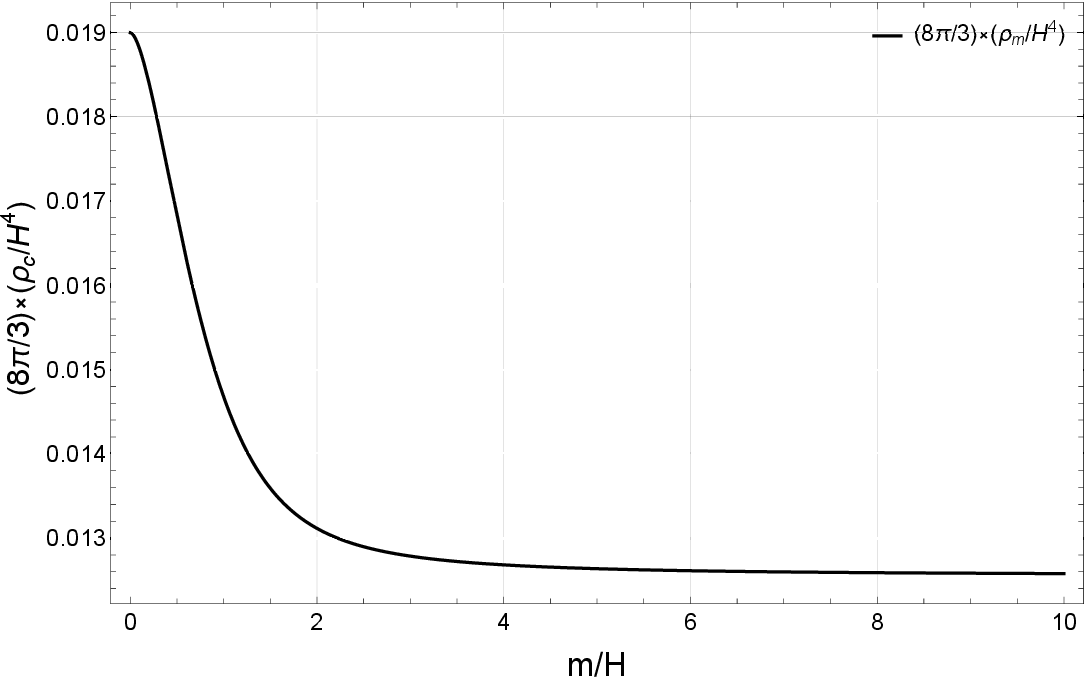}
\caption{Energy density of the minimally coupled scalar field:  $\frac{8\pi G}{3}\rho_m$. }
    \label{Fig2}
\end{figure}

Fig.~\ref{Fig2} shows that the energy density
$\rho_m\equiv\Lambda_m$ is positive definite over the entire range of $(m/H)$,
indicating that \eqref{eqiatop2} can serve as an effective cosmological constant.
Unlike the conformally coupled case shown in Fig.~\ref{Fig1},
the energy density of a minimally coupled scalar field decreases with $(m/H)$.

In the small-mass limit, Eq.~\eqref{eqiatop2} reduces to
\bl
\lim_{m^2/H^2\to 0}\Lambda_m
\approx 3\times\frac{H^4}{16\pi^2},
\label{masslesslimit}
\el
which scales as $H^4$ and is independent of the mass.
In the large-mass limit, Eq.~\eqref{eqiatop2} becomes
\bl
\lim_{m^2/H^2\to\infty}\Lambda_m
\approx \frac{119}{60}\times\frac{H^4}{16\pi^2},
\label{massivelimit}
\el
which also scales as $H^4$ and is independent of the mass.
The two limits differ only by a numerical factor of order unity.
This indicates that the effective cosmological constant generated by
a minimally coupled scalar field is not sensitive to the ratio $(m/H)$.
Over the entire mass range, it remains of the same order of magnitude,
as illustrated in Fig.~\ref{Fig2}.

Treating the minimally coupled scalar field as the sole source of the Friedmann equation, one has
\bl
\Big(\frac{a'}{a}\Big)^2
=\frac{8\pi G a^2}{3}\rho_m.
\label{eqmr}
\el
Using the above results, Eq.~\eqref{eqmr} can be written as
\bl
H^2
=\frac{8\pi G}{3}\,
C_m\!\Big(\frac{m^2}{H^2}\Big)
\frac{H^4}{16\pi^2},
\label{eqjio}
\el
where $C_m$ is a function of $m^2/H^2$ and takes values in the range
$(119/60,\,3)$. Eq.~\eqref{eqjio} then leads to
\bl
\frac{1}{G H^2}
=\frac{C_m}{6\pi}
\in\Big(\frac{119}{360\pi},\,\frac{1}{2\pi}\Big)
\sim 10^{-1}.
\label{huiehgiuh}
\el
The corresponding Hubble rate is
\bl
H \approx 10^{1/2} M_{\mathrm{pl}}
\approx 10^{19}\,\mathrm{GeV}.
\label{Hmini}
\el
This value is far above the present Hubble scale
$H_0\approx 10^{-42}\,\mathrm{GeV}$.
Therefore, regardless of its mass, a minimally coupled scalar field cannot
account for the current dark energy.
It also cannot serve as the inflaton, since the required inflationary scale
$H_{\mathrm{Inf}}\approx 10^{14}\,\mathrm{GeV}$ is much smaller than the value implied by Eq.~\eqref{Hmini}.
This conclusion is consistent with Fig.~\ref{Fig2}, where the maximal value of
$\tfrac{8\pi G}{3}\rho_m$ is about $0.019$, far below
$1/(G H_{\mathrm{Inf}}^2)\approx 10^{10}$
and
$1/(G H_0^2)\approx 10^{122}$.

\section{Summary}

In this paper, we examine whether the Friedmann equation in de Sitter space allows consistent solutions during both the inflationary epoch and the present accelerating expansion when the source is a conformally or minimally coupled scalar field. We then assess whether such fields can act as the inflaton or as the present dark energy.

We find that the regularized energy density of a conformally coupled scalar field cannot act as the inflaton or as dark energy in the small mass regime $(m/H)\ll 1$, since such a ratio cannot be realized during either inflation or the current accelerating expansion. In contrast, in the large mass regime with
$m \approx 10\,M_{\mathrm{pl}} \approx 10^{20}\,\mathrm{GeV}$, the conclusion changes. In this case, the field mass determined from the Friedmann equation is proportional to $G^{-1/2}$ and independent of the Hubble rate. The field thus can act as a candidate for both the inflaton and present dark energy, suggesting that they may share a common quantum origin. Because the field is extremely massive, particle excitations are effectively absent, and it can be probed only through its vacuum effects.

For a minimally coupled scalar field, the regularized energy density is insensitive to the ratio $(m/H)$ and leads to a Hubble rate
$H \approx M_{\mathrm{pl}} \approx 10^{19}\,\mathrm{GeV}$. This value is very different from both the inflationary Hubble scale
$H_{\mathrm{Inf}} \approx 10^{14}\,\mathrm{GeV}$ and the present value
$H_0 \approx 10^{-42}\,\mathrm{GeV}$. Consequently, a minimally coupled scalar field cannot serve as either the inflaton or dark energy, regardless of its mass.

\section{Discussion and Outlook}

The description of inflation adopted here differs from the standard picture. 
In conventional approaches, a scalar field is decomposed into a homogeneous 
background component and perturbations. The background evolves under a suitable 
potential and drives a quasi-exponential expansion, while the perturbations are 
quantized and provide the seeds of the cosmic microwave background anisotropies 
and large-scale structure. In the present work, we instead treat the scalar as a 
single quantum field and ask whether its regularized vacuum stress tensor can 
itself support an inflationary de Sitter phase. Within the fixed-background 
approximation, the Friedmann equation indicates that a conformally coupled scalar 
field can provide such a source. The regularized vacuum stress tensors \eqref{cctmunu} and \eqref{mctmunu} have 
been obtained in exact de Sitter space, where the Hubble parameter is constant. 
This setting allows the contribution of the vacuum stress tensor to be isolated 
in a controlled way. A complete cosmological scenario, however, requires going 
beyond this approximation and including the dynamical response of the geometry. 
In particular, backreaction effects may lead to a departure from the exact 
de Sitter phase and provide a possible mechanism for ending inflation.

Several directions remain open. First, it would be useful to clarify under what 
conditions a highly symmetric de Sitter initial state can be dynamically selected 
or approximated at very high energy scales. Second, the exit from inflation and 
the subsequent reheating stage should be studied, possibly through backreaction 
or through couplings between the scalar field and other degrees of freedom that 
convert vacuum energy into particle excitations. Third, a phenomenological 
analysis is needed to determine whether this framework can reproduce current 
observational constraints, including the scalar spectral index and the 
tensor-to-scalar ratio \cite{KalloshLinde2025}.

For the dark-energy application, the isolated de Sitter analysis should also be 
extended. Two issues are particularly relevant. The first is whether a 
quasi-de Sitter phase can emerge naturally after the dust-dominated era. The 
second is whether the backreaction of the regularized stress tensor of a 
conformally coupled scalar field can generate an effective equation of state 
that evolves from the phantom regime to the quintessence regime, as suggested by 
recent DESI data \cite{DESI}.

\section*{Acknowledgements}

The author thanks Yang Zhang for his insightful discussions.


\begin{thebibliography}{99}






\bibitem{Liddle1993} A. R. Liddle, Phys. Rev. D 29 (1994) 2.


\bibitem{Planck2018}
N. Aghanim, Y. Akrami, M. Ashdown, et al.,
Astron. Astrophys. 641 (2020) A6.

\bibitem{Weinberg2008}
S. Weinberg, \textit{Cosmology}, (Oxford University Press, Oxford, 2008).

\bibitem{Guth1981} A.~H. Guth, Phys. Rev. D 23 (1981) 347.

\bibitem{Lindde1982} A.D. Linde, Phys. Lett. B108 (1982) 389.

\bibitem{KalloshLinde2025}
R.~Kallosh, A.~Linde,
Gen. Relativ. Gravit. 57 (2025) 135.

\bibitem{Bamba024}
K. Bamba, Universe, 10(3) (2024).


\bibitem{Riess1998} A.~G. Riess, A.~V. Filippenko, P. Challis, et al., Astron. J. 116 (1998) 1009--1038. 

\bibitem{Perlmutter1999} S. Perlmutter, G. Aldering, D. Goldhaber, et al., Astrophys. J. 517 (1999) 565--586. 

\bibitem{Suzuki2012} N. Suzuki, D. Rubin, C. Lidman, et al., Astrophys. J. 746 (2012) 85.

\bibitem{Hinshaw2013} G. Hinshaw, D. Larson, E. Komatsu, et al., Astrophys. J. Suppl. Ser. 208 (2013) 19.

\bibitem{Alam2021} S. Alam, M. Aubert, S. Avila, et al., Phys. Rev. D 103 (2021) 083533.

\bibitem{Starobinsky1980} A.~A. Starobinsky, Phys. Lett. B 91 (1980) 99.

\bibitem{Bezrukov2008} F.~L. Bezrukov, M. Shaposhnikov,  Phys. Lett. B 659 (2008) 703.

\bibitem{KalloshLinde2013} R. Kallosh, A. Linde, JCAP 07 (2013) 002.

\bibitem{Caldwell1998} R.~R. Caldwell, R. Dave, P.~J. Steinhardt, Phys. Rev. Lett. 80 (1998) 1582. 

\bibitem{Caldwell2002} R.~R. Caldwell, Phys. Lett. B 545 (2002) 23. 

\bibitem{ArmendarizPicon2001} C. Armendariz-Picon, V. Mukhanov, P.~J. Steinhardt, Phys. Rev. D 63 (2001) 103510.

\bibitem{Zheng2022}
J. Zheng, S. Cao, Y.~J. Lian, et al.,
Eur. Phys. J. C 82 (2022) 582.

\bibitem{Zhang2007}
Y. Zhang, T.~Y. Xia, W. Zhao,
Class. Quantum Grav. 24 (2007) 3309--3337.

\bibitem{Xia2007}
T.~Y. Xia, Y. Zhang,
Phys. Lett. B 656 (2007) 19--24.

\bibitem{Wang2008a}
S. Wang, Y. Zhang, T.~Y. Xia,
JCAP  10 (2008) 037.

\bibitem{Wang2008b}
S. Wang, Y. Zhang,
Phys. Lett. B 669 (2008) 201.

\bibitem{Zhao2009}
W. Zhao,
Int. J. Mod. Phys. D 18 (2009) 1331.

\bibitem{Don2016}
P. Don, A. Marcianò, Y. Zhang, et al.,
Phys. Rev. D 93 (2016) 043012.

\bibitem{Bose2025} WeiNan Xiao, Xuan Ye, Yang Zhang, and A. Marciano j.ucas, 10.7523 (2024)  029.

\bibitem{Carroll2004}
S.~M.~Carroll, V.~Duvvuri, M.~Trodden and M.~S.~Turner,
Phys. Rev. D 70  (2004) 043528.



\bibitem{Nicolis2009}
A.~Nicolis, R.~Rattazzi and E.~Trincherini,
Phys. Rev. D 79  (2009) 064036.



\bibitem{deRham2011}
C.~de Rham, G.~Gabadadze and A.~J.~Tolley,
Phys. Rev. Lett 106 (2011) 231101.



\bibitem{Bengochea2009}
G.~R.~Bengochea and R.~Ferraro,
Phys. Rev. D 79  (2009) 124019.

\bibitem{Maggiore2018}
M. Maggiore, \textit{Gravitational Waves, Volume 2: Astrophysics and Cosmology}, (Oxford University Press, Oxford, 2018).

\bibitem{Mukhanov2005}
V. Mukhanov, \textit{Physical Foundations of Cosmology}, (Cambridge University Press, Cambridge, 2005).


\bibitem{LiddleLyth2000}
A. R. Liddle, D. H. Lyth, \textit{Cosmological Inflation and Large-Scale Structure}, (Cambridge University Press, Cambridge, 2000).


\bibitem{Dodelson2020} S. Dodelson, F. Schmidt, \textit{Modern Cosmology}, 2nd ed., (Academic Press, Cambridge, MA, 2020).

\bibitem{weinberg1989} S. Weinberg, Rev. Mod. Phys. 61: (1)
(1989) 1.

























\bibitem{Kaneta} 
K. Kaneta, K. Oda, M. Yoshimura,
arXiv:2503.02409 (2025).

\bibitem{Joan} J. Sola, Journal of Physics A, 41 (2008) 16.

\bibitem{Rubio} J. Rubio and C. Wetterich,
Phys. Rev. D 96 (2017) 063509.





\bibitem{ZhangYe2025b}
Y.~Zhang and X.~Ye,
Universe 11 (2), 72 (2025).

\bibitem{YeZhang2025}
X.~Ye and Y.~Zhang,
arXiv:2509.23388 (2025).


\bibitem{YeZhang2024}
X.~Ye and Y.~Zhang,
Int. J. Mod. Phys. D 33, 2450050 (2024).

\bibitem{ZhangYe2022}
Y.~Zhang and X.~Ye,
Phys. Rev. D 106, 065004 (2022).

\bibitem{XuanJCAP2022} Xuan Ye, Yang Zhang, and Bo Wang, J. Cosmol. Astropart. Phys. 09 (2022) 020.

\bibitem{ZhangYeWang2020}
Y.~Zhang, X.~Ye and B.~Wang,
Sci. China Phys. Mech. Astron. 63, 250411 (2020).

\bibitem{ZhangWangYe2020}
Y.~Zhang, B.~Wang and X.~Ye,
Chin. Phys. C 44, 095104 (2020).

\bibitem{ParkerFulling1974}
L.~Parker and S.~A.~Fulling,
Phys.\ Rev.\ D 9, 341 (1974).

\bibitem{Ye2023}
X.~Ye, 
\textit{The regularization of the quantum fields in curved spacetime---The quantum origin of cosmological constant} (in Chinese), 
Ph.D. thesis, University of Science and Technology of China (2023).


\bibitem{DESI} DESI Collaboration, arXiv: 2503.14745 (2025).


\end{thebibliography}
\end{document}